\documentclass[superscriptaddress, twocolumn]{revtex4}
\usepackage{graphicx}
\usepackage{newfloat}
\DeclareFloatingEnvironment[name={SUPP. FIG.}]{suppfigure}
\newcommand{\unit}[1]{\ensuremath{\;\mathrm{#1}}}
\newcommand{\ket}[1]{\left| #1 \right\rangle}

\newcommand{\uibk}{Institut f\"ur Experimentalphysik, Universit\"at Innsbruck, Technikerstr. 25, 6020 Innsbruck, Austria}
\newcommand{\iqc}{Institute for Quantum Computing, University of Waterloo, 200 University Ave. West, Waterloo, Ontario, Canada}
\newcommand{\win}{Waterloo Institute for Nanotechnology, University of Waterloo, 200 University Ave. West, Waterloo, Ontario, Canada}
\newcommand{\nrc}{National Research Council of Canada, 1200 Montreal Road, Ottawa, Ontario, Canada}

\begin{document}

\title{ Polarization entangled photons from quantum dots embedded in nanowires}

\author{Tobias Huber}\email{tobias.huber@uibk.ac.at}\affiliation{\uibk}
\author{Ana Predojevi\'{c}}\affiliation{\uibk}
\author{Milad Khoshnegar}\email{m3khoshn@uwaterloo.ca}
\affiliation{\iqc}
\affiliation{\win}
\author{Dan Dalacu}\affiliation{\nrc}
\author{Philip J. Poole}\affiliation{\nrc}
\author{Hamed Majedi}
\affiliation{\iqc}
\affiliation{\win}
\author{Gregor Weihs}\affiliation{\uibk}

\maketitle
Linear optical quantum computation~\cite{Knill01} as well as most quantum communication protocols~\cite{Tittel01} require photon entanglement. Additionally, entanglement interconnects the nodes of a quantum network~\cite{Bennett95a} and enables different processor architectures and an increase of computational power~\cite{Ward14}. Quantum dots are promising candidates for generating polarization entangled photon pairs from the biexciton-exciton recombination cascade~\cite{Alpian06}. In contrast to other sources of entangled photon pairs~\cite{Predojevic12}, in quantum dots it is possible to create photon pairs deterministically in a coherent way~\cite{Jayakumar13} and they posses inherent sub-poissonian photon statistics~\cite{Predojevic14}. We present entanglement generated from a novel structure: a single InAsP quantum dot embedded in an InP nanowire. These structures can grow in a site controlled way and exhibit high collection efficiency; we detect 0.5 million biexciton counts per second coupled into a single mode fiber. For the entanglement we observe a fidelity of 0.76(2) to a reference maximally entangled state as well as a concurrence of 0.57(6).

Nanowire (NW) quantum dots were proposed to deliver superior performance as entangled photon pair sources because of their high symmetry~\cite{Singh09} and considerably enhanced light extraction efficiencies of axial excitation and collection~\cite{Reimer12}. Although a lot of experimental efforts focused on the generation of entangled photon pairs from semiconductor quantum dots~\cite{Stevenson06,Dousse10,Jayakumar14}, we are not aware of any report on the experimental realization of entangled photon pairs from NW quantum dots. Furthermore, NW quantum dots offer some unique features in addition to those already explored in Stranski-Krastanov (SK) grown quantum dots. The versatility in the axial and radial growth of III-V nanowires~\cite{Bjork04} allows to manipulate both their electronic and optical properties. The excitonic energy states of NW quantum dots can be deterministically modified by controlling the growth parameters. Finally, single quantum dots can be easily stacked up in a nanowire~\cite{Bjork04}, forming quantum dot molecules with unprecedented design flexibility.

\begin{figure}[h]
\centerline{\includegraphics[width=1\columnwidth]{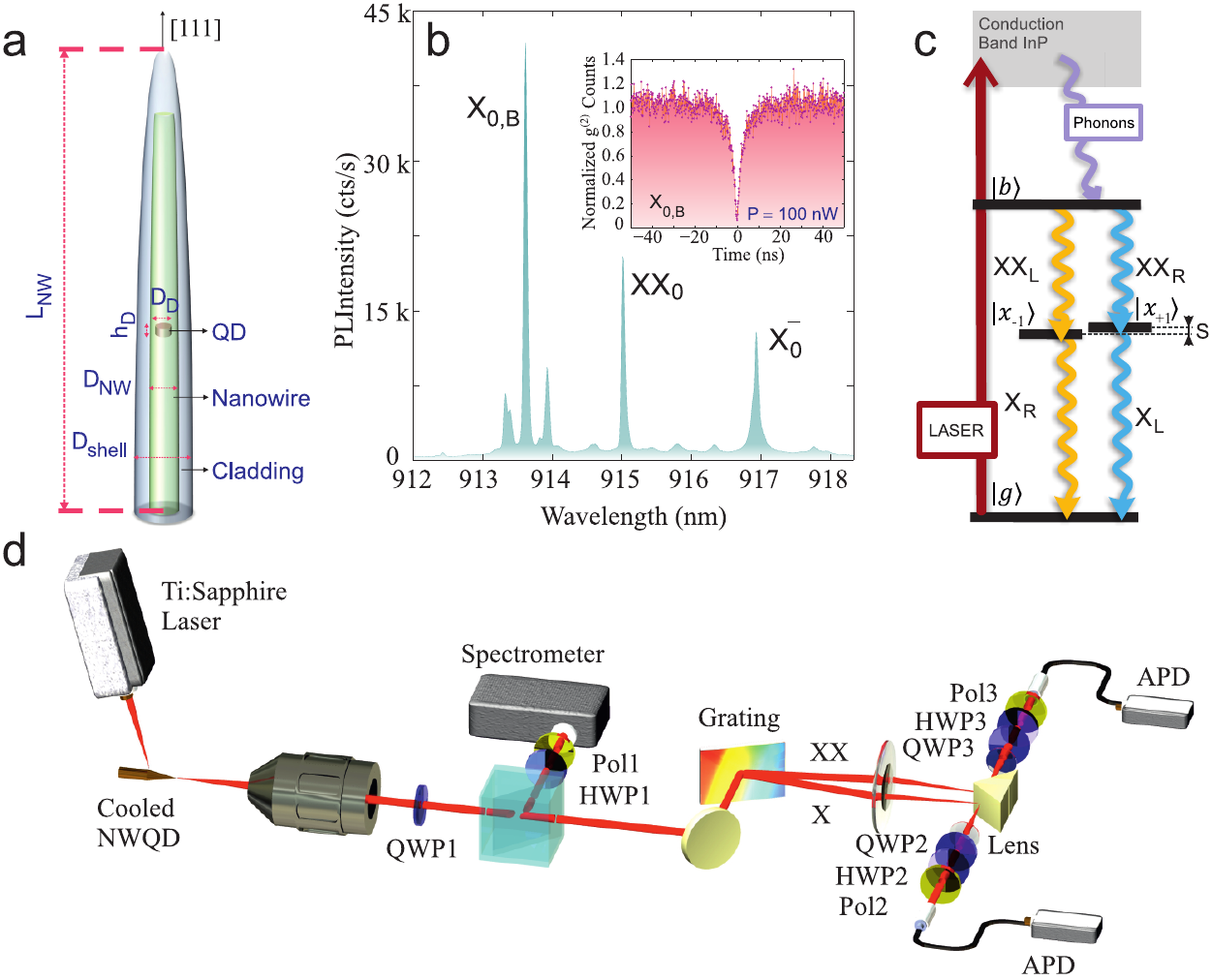}}
\caption{ \label{fig:setup} Schematic, spectrum, and energy scheme of the NW quantum dot and experimental setup.
\textbf{a} Schematic of a clad NW quantum dot tapered at the top. $h_\mathrm{D}$: quantum dot height $\sim 6-8 \unit{nm}$; $D_{\mathrm{D}}$: quantum dot diameter $28\unit{nm}$; $D_{\mathrm{NW}}$: nanowire (core) diameter $\sim 28 \unit{nm}$; $D_{\mathrm{shell}}$: cladding diameter $\sim 200\unit{nm}$; $L_{\mathrm{NW}}$: nanowire length. \textbf{b} Photoluminescence intensity of $s$-shell excitonic resonances of the quantum dot studied in this paper (dot A): exciton $X_{0,\mathrm{B}}$, biexciton $XX_{0}$ and negative trion $X^{-}_{0}$ in continuous-wave (cw) non-resonant excitation with a power of $P_\mathrm{exc}=500\unit{nW}$. The inset illustrates the normalized autocorrelation counts $g^{(2)}$ from the  $X_{0,\mathrm{B}}$ line at low excitation power $P_\mathrm{exc}=100\unit{nW}$, confirming a low multi-photon contribution. Similar $g^{(2)}$-patterns are observed for the other spectral features.
\textbf{c} Energy level structure of the NW quantum dot. The excitation laser creates charge carriers in proximity to the InP band gap. Electrons and holes can relax into the quantum dot by phonon interactions and can fill the biexciton state $\ket{b}$. The recombination can happen on the left or the right path via the exciton spin down or spin up $\ket{x_{-(+)1}}$ state. \textbf{d} Experimental setup. The quantum dot was non resonantly excited using a ps-pulsed Ti:Sapphire laser. The emission was collected using an high NA objective and was analyzed using either a spectrometer and a CCD-camera or using a fiber-coupled dual output grating spectrometer and APDs. Quarter wave plate (QWP) 1 converts the predominant circular polarization of emitted photons into the rectilinear basis. Half wave plate (HWP) 1 and Polarizer (Pol) 1 were used to measure the fine-structure splitting. QWP 2(3), HWP2(3) and Pol 2(3) were used to project the photon states on distinct polarizations to reconstruct the density matrix of the photon pair.
}
\end{figure}

We performed fine-structure splitting (FSS) and correlation measurements on the emission of the NW quantum dots using the experimental setup shown in Fig.~\ref{fig:setup}(d).
Polarization analyzers as depicted in Fig.~\ref{fig:setup}(d) were employed to independently manipulate the polarization state of $X_{0}$ and $XX_{0}$ photons.

\begin{figure}[h]
\centerline{\includegraphics[width=1\columnwidth]{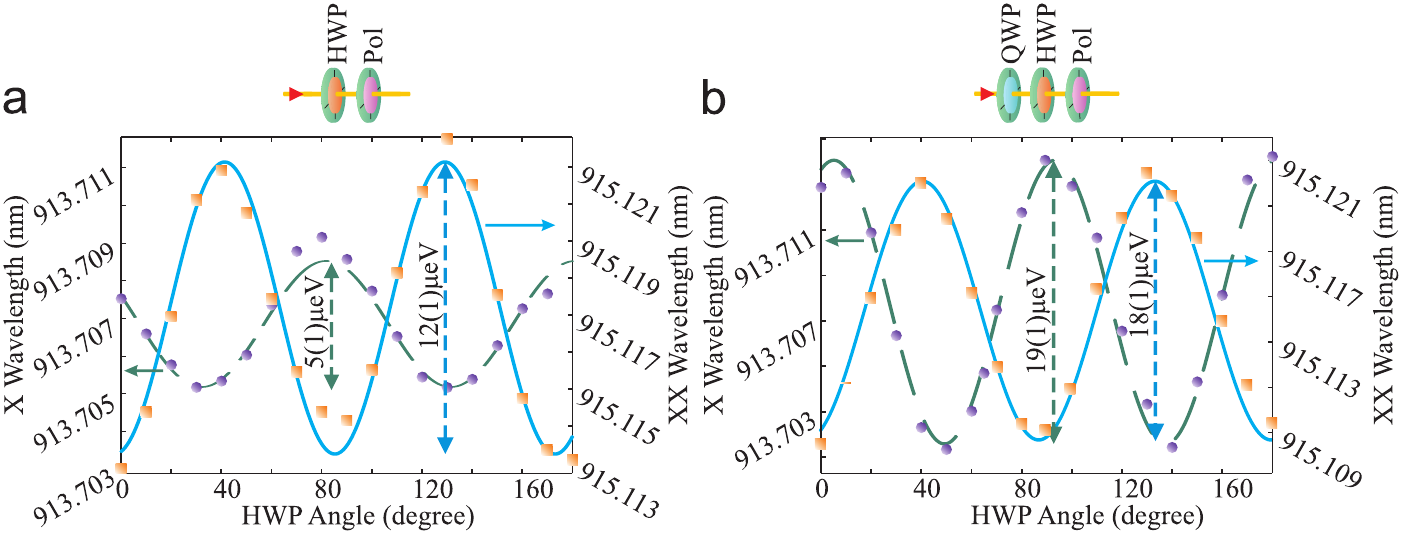}}
\caption{ \label{fig:fss} Fine-structure splitting of dot A. The purple circles (orange squares) and the green dashed (blue solid) fit line represent $X_{0}$ ($XX_{0}$) recombination photons. \textbf{a} shows the evolution of resonances when the emission is guided through HWP1 and Pol1. The $X_{0}$ oscillation differs more than a factor of 2 from the result of the $XX_{0}$ oscillation. We note that since $XX_{0}$ itself exhibits no exchange splitting, its wavelength oscillation must arise exclusively from the FSS of the $X_{0}$ state, i.e. $S_{X_{0}}=S_{XX_{0}}$. \textbf{b} shows the same measurement as in (a) with the additional fixed QWP1 in front of the rotating HWP1. In this case, the amplitudes of $X_{0}$ and $XX_{0}$ oscillations are equal within the error. This measurement shows that the FSS in a system with predominantly circular polarization is strongly underestimated if the basis of emission is not rectilinear.
}
\end{figure}

Despite the fact that the grown NW quantum dots are geometrically symmetric, we noticed a finite level of FSS in practice. We initially measured the  FSS by rotating a HWP in front of a fixed linear polarizer and guiding the quantum dot emission to a grating spectrometer (see Fig.~\ref{fig:setup}(d)). For this measurement the quarter-wave plate 1 (QWP1) was removed. The narrow $X_{0}$ and $XX_{0}$ spectral lines were then fitted to Lorentzian line shapes to resolve the oscillation of the projected states as a result of the half-wave plate (HWP) rotation. The oscillations showed relatively moderate FSSs in the range of 4(1) to 12(2)$\unit{\mu eV}$ for all studied quantum dots in different samples. The $X_{0}$ and $XX_{0}$ oscillations versus the HWP angle is plotted in Fig.~\ref{fig:fss}(a) for dot A. This measurement, however, underestimates the actual FSS because the quantum dot photons couple to the wire's circular polarization as we will show later in the paper. A measurement on a mixture of circularly polarized light presents no significant variation of count rate when analyzed via a rotating linear polarizer.
For this reason we repeated the FSS measurement with QWP1 in the common emission path. The result can be seen in Fig.~\ref{fig:fss}(b) with a value of $S=18(1)\unit{\mu eV}$ for the FSS.
\\

For generating an entangled photon pair, the quantum dot state is prepared as biexciton (see  Fig.~\ref{fig:setup}(c)). After the emission of the biexciton photon, the quantum dots spin state is entangled with the polarization of the biexciton photon. Since the spin up and the spin down exciton states are no eigenstates of the system if they are not degenerate~\cite{Bayer02}, the exciton state will evolve with time. E.g. for the emission of a right circular (R) polarized biexciton photon the exciton will result in the state~\cite{Wilk07}
\begin{equation}
\ket{\Psi_{QD}}=\frac{1}{\sqrt{2}}(e^{i\phi/2}\ket{x_{+1}}-i e^{-i\phi/2}\ket{x_{-1}}).
\label{eq:psi_qd}
\end{equation}
The phase $\phi=S\tau/\hbar$, where $\tau$ is the time elapsed between the first and second photon emission, is directly transfered on the phase of the exciton photon when the exciton recombines. Thus the exciton photon wavefunction is
\begin{equation}
\ket{\Psi_{x-Photon}}=\frac{1}{\sqrt{2}}(e^{i\phi}\ket{L}-i e^{-i\phi}\ket{R}).
\label{eq:psi_x}
\end{equation}
Considering the two-photon state this will lead to an evolution between the $\ket{\Phi^+}=\frac{1}{\sqrt{2}}(\ket{RL}+\ket{LR})$ and $\ket{\Phi^-}=\frac{1}{\sqrt{2}}(\ket{RR}+\ket{LL})$ Bell states.
The state could be rewritten in the H/V (horizontal/vertical) polarization basis as~\cite{ Stevenson08}
\begin{equation}
\ket{\Psi}=\frac{1}{\sqrt{2}}(\ket{HH}+e^{i\phi}\ket{VV}),
\label{eq:psi}
\end{equation}
or in the D/A (diagonal/antidiagonal) polarization basis where we get an oscillation between the $\ket{\Phi^+}=\frac{1}{\sqrt{2}}(\ket{DD}+\ket{AA})$  and the $\ket{\Phi^-}=\frac{1}{\sqrt{2}}(\ket{DA}+\ket{AD})$ state similar to the R/L basis.

This shows that a change of the phase in an entangled state can be directly measured in a time-resolved correlation measurement by observing the correct polarization projection, e.g. projecting both photons to R polarization (projection to RR) should show high coincidence probability for times where the two photon state is in $\ket{\Phi^-}$ and low coincidence probability when the state is in $\ket{\Phi^+}$. The opposite behavior is expected for a projection to DD. If the oscillation is visible in two complementary bases, the third complementary basis (e.g. H/V basis) should show classical correlations as reported in Ref.~\cite{Santori02} which does not show oscillations.

\begin{figure}[h]
\centerline{\includegraphics[width=1\columnwidth]{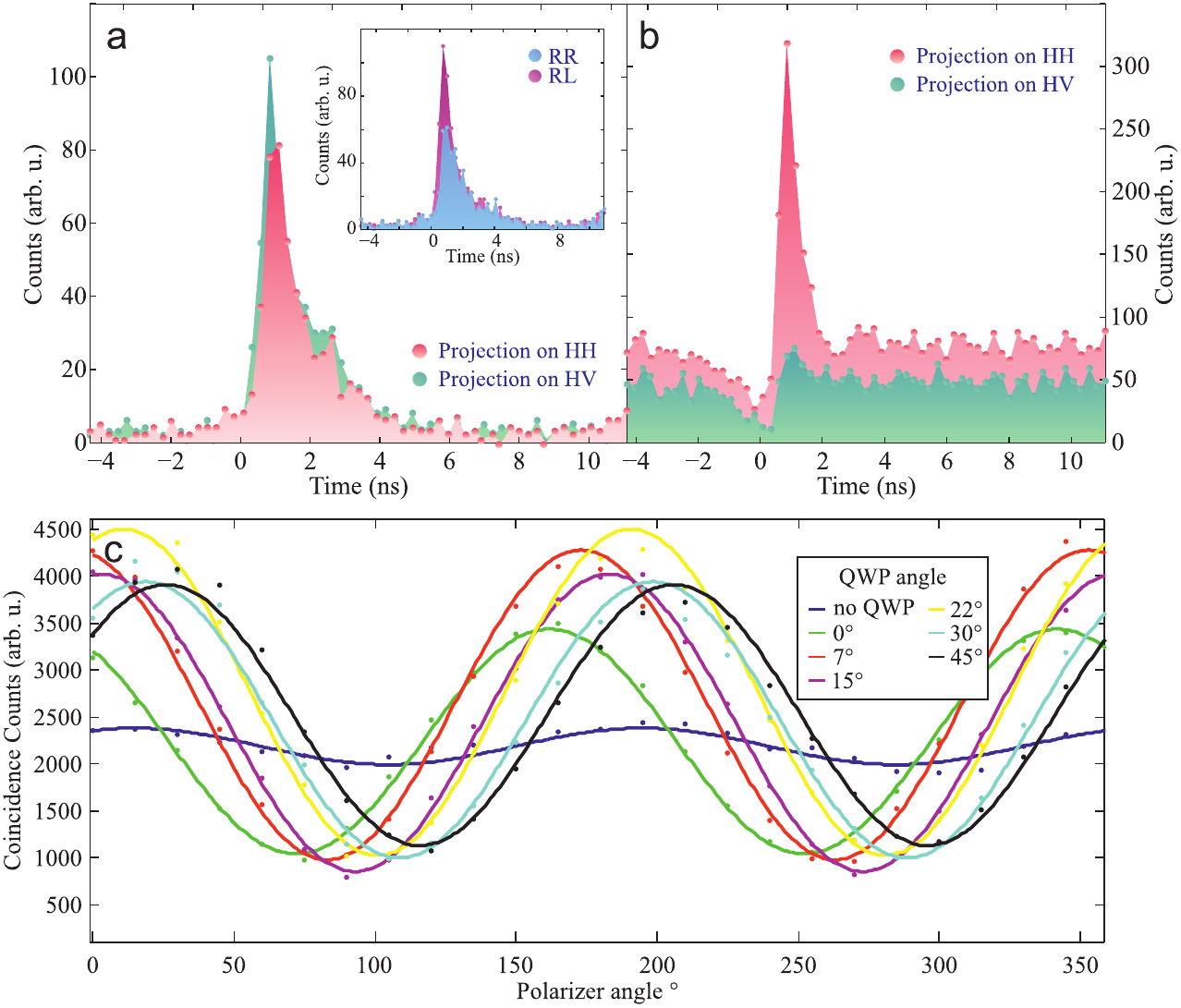}}
\caption{ \label{fig:HH-HV} Comparison of $XX_{0}$-$X_{0}$ cross-correlation coincidences with detectors unable to resolve the time oscillations described in the text. \textbf{a} Coincidence counts after projecting the photons to HH (both photons $XX_{0}$ and $X_{0}$ projected onto horizontal polarization H) (red) and coincidence counts in HV projection (green). Inset: coincidence counts in RR projection (blue) and RL projection (purple), where the projection onto RR basis gives less coincidences than the projection onto RL basis. Comparing the two plots reflects that the quantum dot dipole primarily couples to the nanowires circular basis as the projections onto HH and HV present no significant difference. \textbf{b} Coincidence counts in HH projection (red) and HV projection (green) after inserting QWP1 into the common emission path. Coincidences in HV basis are strongly suppressed. A high level of background counts appears here on account of cw pumping. \textbf{c} Coincidence counts for different QWP1 settings. The exciton was projected to H and the polarizer in front of the biexciton was turned. $0^\circ$ corresponds to a projection on HH and $90^\circ$ corresponds to a projection to HV. }
\end{figure}

For the sake of comparison to previously reported measurements from entangled photon pairs from quantum dots, it would be beneficial if the basis where classical correlations can be observed is H/V. Unfortunately, in our system this is not the case (see Fig.~\ref{fig:HH-HV}(a)). The figure shows almost equal probabilities for HH and HV, respectively. Classical correlations can be partially observed in the R/L basis (see inset in Fig.~\ref{fig:HH-HV}(a)). To get classical correlations in H/V basis, we locally rotate the polarization state by inserting a quarter wave plate (QWP1) into the common emission path of both $XX_0$ and $X_0$ photons (see Fig.~\ref{fig:setup}(d)). For achieving maximum visibility in H/V basis, we performed several coincidence measurements with different QWP1 settings. For each visibility measurement the polarizer in the $X_0$ photon path was fixed to H and the polarizer in the $XX_0$ photon path was rotated (see Fig.~\ref{fig:HH-HV}(c)). Such a local rotation on the polarization qubit cannot affect the degree of entanglement but can change the fidelity to a particular desired state.

To quantify the degree of entanglement, a tomographic experiment was implemented with 16 cross-correlation measurements on different combinations of polarization projections~\cite{James01}. The NW quantum dots were excited by a pulsed laser to minimize uncorrelated photons caused by re-excitations in a continuous pumping scheme. Owing to the nonzero FSS, the evolving phase of the photon pair state reduces the time-integrated concurrence~\cite{Stevenson08}. We therefore postselect the correlated photons within specific time windows and calculate the fidelity with respect to a reference Bell state for each time interval. Accordingly, a temporal resolution sufficient to resolve the correlation pattern of each window is necessary, as the period of the  photon pair phase $\phi$ is predicted to be $\sim 230(12)\unit{ps}$ for an FSS of $S \sim 18(1)\unit{\mu eV}$ (see Fig.~\ref{fig:fss}(b)), i.e. every $\sim 115\unit{ps}$ the phase of photon pair flips while its power simultaneously fades according to the finite exciton lifetime ($\sim 2  \unit{ns}$). In order to resolve the oscillations we used highly time-resolving detectors after the spectral filtering and the polarization state projection.

\begin{figure}[h]
\centerline{\includegraphics[width=1\columnwidth]{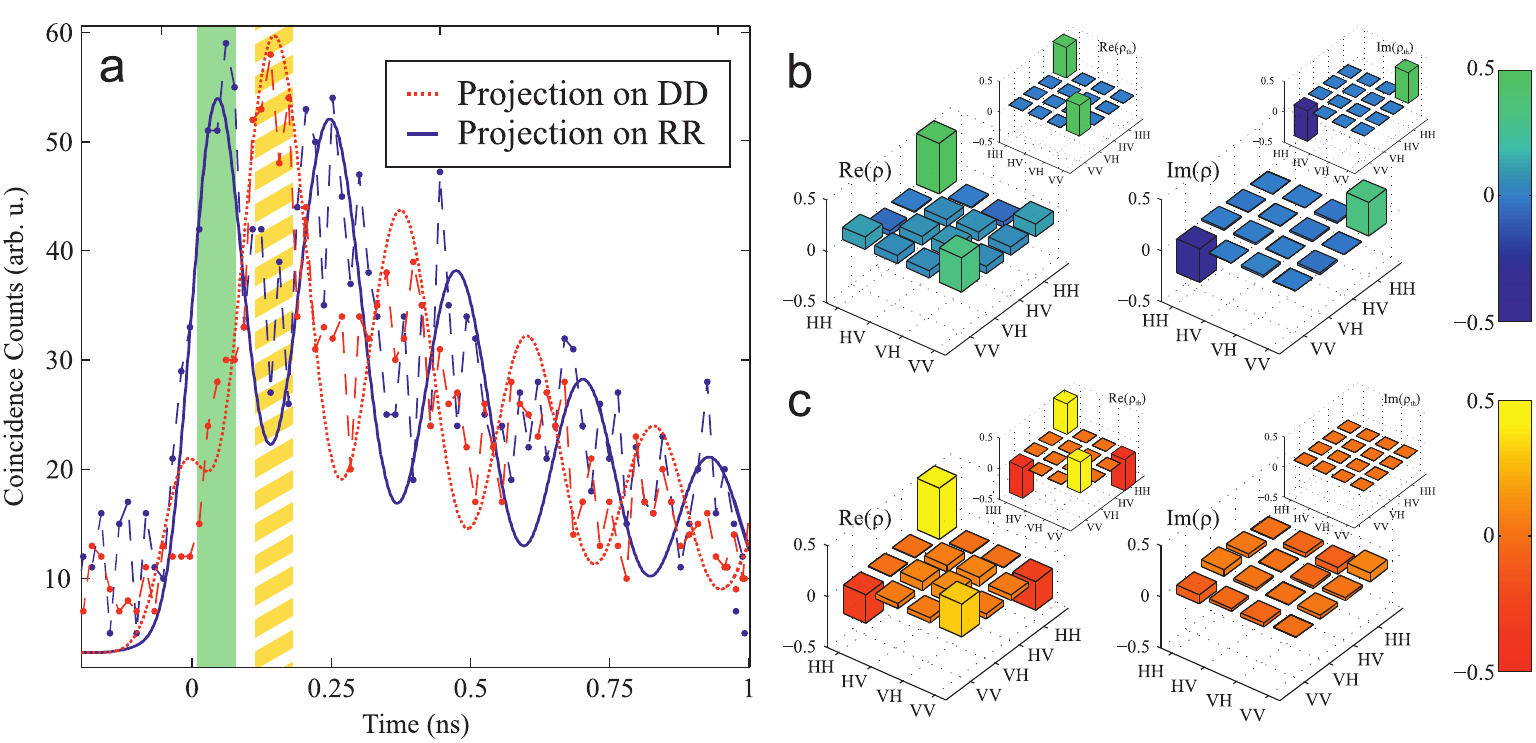}}
\caption{ \label{fig:density_matrix} Tomographic measurement of polarization entanglement. (a) shows two specific analyzer settings while (b,c) shows real and imaginary parts of the density matrices associated with the reconstructed and the corresponding theoretical states.
\textbf{a} Cross-correlation counts in DD (red) and RR (blue) bases resolved with $35\unit{ps}$ temporal resolution. Please note that the separation between the obtained data points is only $16\unit{ps}$. The detectors are still not able to resolve the oscillations fully otherwise their minima would reach zero. The dotted (red) and solid (blue) curves represent the fits to the correlation functions with oscillations of the same period and amplitude and $180\unit{^{\circ}}$ phase difference, indicating the evolving phase of the photon pair state. The dotted curve reaches its local maxima where the blue curve has its local minima and vice versa. The green (solid) and yellow (striped) shaded areas signify the first and second temporal windows where the local maximum (minimum) occurs. \textbf{b} shows the reconstructed density matrix with photons postselected in a time window of $65\unit{ps}$ (green-shaded region in (a)) immediately after the cascade decay. Insets show the density matrix components of the ideal theoretical state. \textbf{c} Similar to (b), but measured at the second time interval (orange-shaded region in (a)), once the photon pair phase has evolved.
}
\end{figure}

Fig.~\ref{fig:density_matrix}(a) shows the cross-correlation counts in both the RR and DD bases obtained with 35 ps temporal resolution. Here, the oscillatory behaviour of photon pair wave function is clearly observable in the corresponding correlation patterns, particularly within the early stages after the cascade initialization. The period of the fitted oscillatory functions $\tau_\mathrm{fit}=225(5)\unit{ps}$ agrees with the value of $230(12)\unit{ps}$ calculated from the FSS value.

In order to identify the photon pair states, the $65\unit{ps}$-wide shaded areas in Fig.~\ref{fig:density_matrix}(a) were postselected and the corresponding density matrices $\rho$ were reconstructed using a maximum-likelihood estimation~\cite{James01} from the raw data without any background subtraction. The reconstructed density matrix $\rho_1$ for the green shaded area is shown in Fig.~\ref{fig:density_matrix}(b). The inset shows the density matrix $\rho_{\mathrm{th1}}$ of the state $\ket{\Psi_{\mathrm{th1}}}=\frac{1}{\sqrt{2}}(\ket{HH}+i\ket{VV})$ to which $\rho_1$ has a fidelity of $F=0.74(2)$. Since we performed a tomographic measurement, we calculated the fidelity directly from $\rho_1$,  $F=\mathrm{Tr}(\sqrt{\sqrt{\rho_1}\cdot\rho_{th}\cdot\sqrt{\rho_1}})^2$, instead of using the correlation visibility in different polarization bases~\cite{Hudson07}. The concurrence of $\rho_1$ is $C=0.57(6)$. Within the following time window, highlighted by the orange-shaded region in Fig.~\ref{fig:density_matrix}(a), a phase rotation occurs as could be inferred from the associated density matrix shown in Fig.~\ref{fig:density_matrix}(c). The fidelity with the reference state $\ket{\Psi_{\mathrm{R}}}=\frac{1}{\sqrt{2}}(\ket{HH}-\ket{VV})$ and the photon concurrence were calculated to be $F=0.69(2)$ and $C=0.45(2)$. In the subsequent window (the next maximum of the RR projection) , the concurrence is zero and the fidelity with the rotated state is 0.52(1), suggesting that the entanglement is lost after $\sim260\unit{ps}$, most probably due to decoherence.

Having a closer look at the fidelities (see Table~\ref{tab:fidelities}) we find that the state is not only represented in the  $\ket{\phi^\pm}$ subspace. The highest fidelity we get to a state which is rotated, for both analyzed coincidence windows. We attribute this to a non perfect state rotation which origins in the coupling of the quantum dot to the nanowires circular polarization. The resulting emission might be not perfectly circular but elliptical to some degree. Considering an elliptical photon polarization the QWP1 will not turn the polarization of the emission to rectilinear basis, introducing an error in the expected photon state. To rule out the possibility that the polarization rotation was introduced by our setup, we tested our setup with a polarized laser beam and a polarimeter.

\begin{table}[h]
\caption{Fidelities of the reconstructed density matrices to different maximally entangled states.}
\label{tab:fidelities}
\begin{tabular}{ l | c | r }
Reference state & F to $\rho_1$ & F to $\rho_2$ \\
\hline
\hline
$\ket{\phi^+}=\frac{1}{2}(\ket{HH}+\ket{VV})$ 	& 0.53(2) & 0.13(2) \\
$\frac{1}{2}(\ket{HH}+i\ket{VV})$ 			& 0.74(2) & 0.50(2) \\
$\ket{\phi^-}=\frac{1}{2}(\ket{HH}-\ket{VV})$ 	& 0.29(2) & 0.69(2) \\
$\frac{1}{2}(\ket{HH}-i\ket{VV})$ 			& 0.08(1) & 0.33(2) \\
$\frac{1}{2}(\ket{HH}+e^{i\cdot70^\circ}\ket{VV})$ & 0.76(2) &  \\
$\frac{1}{2}(\ket{HH}+e^{i\cdot160^\circ}\ket{VV})$ 	&  & 0.70(2) \\
\end{tabular}
\end{table}

We utilized postselection of quantum correlated photons before the photon pair coherence is lost. We observed rather high levels of fidelity, $F=0.76(2)$ ($C=0.57(6)$) and $F=0.70(2)$ ($C=0.45(2)$), despite the comparably fast phase variation of the photon pair state. The periodicity of the phase oscillation was first estimated directly via FSS measurements, then validated by the oscillatory behaviour of the cross-correlation patterns. Our results serve as a prototypical assessment of the optical quality of NW quantum dots for generating polarization entangled photon pairs.

\section*{Methods}

The NW quantum dots investigated here are ternary In(As)P insertions embedded inside [111]-oriented tapered InP nanowires grown in the wurtzite phase surrounded by a cladding (see Fig.~\ref{fig:setup}(a)). The radial growth of clad InP nanowires provides the option to increase their thickness for optical confinement and allows for coupling between the quantum dot dipole and the nanowire guided modes. From the dispersion diagram of the guided modes in a cylindrical waveguide we can infer that beyond the fundamental mode, the second mode that the quantum dot can couple to emerges for $D_{\mathrm{D}}> 0.23\lambda_{0}$~\cite{Bleuse11}. Here, $D_{\mathrm{D}}$ stands for the nanowire diameter and $\lambda_{0}$ is the mode wavelength in free space. To achieve the highest extraction efficiency the optimal range of cladding diameters has been numerically demonstrated to be $0.2<D_{\mathrm{D}}/\lambda_{0}<0.25$ \cite{Friedler09, Reimer12}. The $s$-shell excitonic resonances of the quantum dots under study were tuned to be around $910\text{ - }920\unit{nm}$ (see Fig.~\ref{fig:setup}(b)), hence the cladding diameter is set to $\sim 200\unit{nm}$ ($D_{\mathrm{D}}<0.23\times 910\unit{nm}$) to avoid multi-mode coupling.

The axial aspect ratio $a_{h}$ of NW quantum dots, defined as the ratio between their height $h_{\mathrm{D}}$ and diameter, determines the strength of the single particle localization and tailors the dispersion of the hole. In order to save the oscillator strength against the internal [111]-oriented piezoelectric field~\cite{Schliwa09} and the exchange-induced spin-flip and cross-dephasing processes~\cite{Tsitsishvili05}, adequately strong axial quantization is favorable ($a_{h}<0.3$)~\cite{Khoshnegar12}. In addition, the single particle orbitals are very small in very flat quantum dots ($a_{h}<0.15$), and they may present larger anisotropic exchange splitting under morphological asymmetries. A moderate axial localization ($0.15<a_{h}<0.3$) maintains the hole ground state $h_{0}$ predominantly heavy hole ($hh$)-like~\cite{Niquet2008}, i.e. $\ket{J^{h_{0}},J^{h_{0}}_{z}}=\ket{3/2,\pm3/2}$ ($J^{h_{0}}$ represents the total angular momentum of the $s$-shell hole), thus the quantum dot dipole effectively couples only to circularly polarized light. That the NW quantum dot emission is mostly circularly polarized if collected along the NW axis was also found in Ref.~\cite{vanWeert09a}.

The nanowires are grown on a (111)B semi-insulating Fe-doped InP substrate coated with a $20\unit{nm~SiO_{2}}$ layer, which is electron-beam patterned with arrays of holes to define the gold catalyst size. Gold is then deposited in the $\mathrm{SiO_{2}}$ holes via a self-aligned lift-off process, which allows the nanowires to grow at determined locations on the substrate. The growth of clad nanowires includes two primary phases: Phase I: The nanowire core region is grown employing chemical beam epitaxy (CBE) with trimethylindium (TMI) and precracked $\mathrm{PH_{3}}$ and $\mathrm{AsH_{3}}$ in the vapor-liquid-solid (VLS) mode at $420\unit{^{\circ}C}$~\cite{Dalacu11}. Stabilizing the temperature at $420\unit{^{\circ}C}$ ensures that InP only grows vertically at the interface of InP and the gold catalyst and prevents any radial growth. The quantum dot diameter is determined by the size of the gold catalyst. Subsequently, the quantum dot height is decided by the time interval that $\mathrm{AsH_{3}}$ replaces $\mathrm{PH_{3}}$ (2 - 3 s). Defect-free wurtzite nanowires are obtained upon TMI fluxes below a definite threshold value which is dependent on the group V flux~\cite{Poole12}. The total growth time of the nanowire core is 26 min. Phase II: Radial growth of the cladding is accomplished by threefold increase of the $\mathrm{PH_{3}}$ flow which significantly suppresses the axial growth at the nanowire tip and promotes non-catalyzed radial growth~\cite{Dalacu12}. The radial growth time is almost 4 times the catalyzed growth time ($\sim 96 \unit{min}$), and the final height of the clad nanowire, i.e. $L_\mathrm{NW}$ in Fig. \ref{fig:setup}(a), reaches approximately twice the height of its core.

 The sample was held at 5 K in a temperature-stabilized liquid flow cryostat. The quantum dots were pumped non-resonantly ($\lambda_{\mathrm{exc}}=835 \unit{nm}$ using a Ti:Sapphire laser in cw or ps-pulsed mode), above the donor-acceptor recombination level usually observed at $\sim 1.44 \unit{eV}$ and in proximity to the wurtzite InP band gap $\sim1.50 \unit{meV}$
, to photogenerate carriers in the nanowire continuum. The excitation beam was focused on the spatially isolated nanowire from the side. The quantum dot photoluminescence was collected by an objective lens with a numerical aperture (NA) of 0.7 and dispersed by grating spectrometers (spectral resolution $\sim0.01 \unit{nm}$) in order to separate the exciton $X_{0}$ and biexciton $XX_{0}$ photons and send them to different polarization analyzers and photon detectors.
We applied avalanche photodiodes (APD) with $\sim 300\unit{ps}$ temporal resolution for correlation experiments not requiring a high time resolution, and fiber-pigtailed single-photon detection modules from Micro Photon Devices (MPD) with a high temporal resolution of $35\unit{ps}$ for polarization entanglement measurements. The high resolution detectors had a quantum efficiency below $5\unit{\%}$ at wavelengths above $900\unit{nm}$ which leads to integration times of $20\unit{min}$ per projection to resolve the cross-correlation pattern. A time-tagging module was utilized to record correlations between the detection modules. The typical count rate of the MPD detectors during the correlation measurements was $\sim65 \unit{KHz}$ for $X_{0}$ and $\sim23 \unit{KHz}$ for $XX_{0}$ after polarization projection.



\section{Supplementary Information}

\section*{Origin of the fine-structure splitting} 

An anisotropic FSS results from the exchange energy caused by the Rashba spin-orbit interaction in combination with a low symmetry of electron and hole orbitals. The anisotropic exchange splitting is theoretically predicted to vanish once the net symmetry of the exciton wave-function exceeds $C_{2v}$~\cite{Singh09, Altmann94}. The ideal hexagonal ($D_{6h}$) or cylindrical ($D_{\infty h}$) symmetry of [111]-oriented zinc blende (or wurtzite) NW quantum dots leads to $C_{3v}$-symmetric orbitals, which preserve the degeneracy of the bright excitons $X_{0,\mathrm{B}}$. This elevated symmetry character directly originates from the $C_{3v}$ symmetry of the strain-induced potentials, including the piezoelectric potential. The origin of the FSS in our NW quantum dots could be explained either by a) an in-plane (perpendicular to the nanowire axis) asymmetry, namely elongation, b) off-center growth of the nanowire core, or c) inhomogeneity of the quantum dot material composition, which all may bring down the symmetry of the net confinement and lift the $X_{0,\mathrm{B}}$ degeneracy. In the following, we discuss each of above potential sources of FSS.

\begin{suppfigure}[h]
\centerline{\includegraphics[width=1\columnwidth]{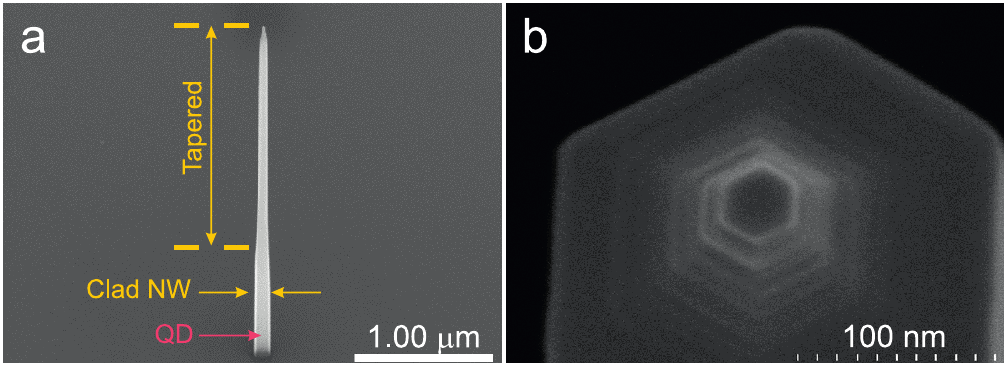}}
\caption{Scanning electron microscopy (SEM) images of the nanowires. \textbf{a} SEM image of a single tapered nanowire, which is spatially isolated for convenient optical access. \textbf{b} Top view SEM image of a clad nanowire showing the in-plane hexagonal symmetry of its core (and the embedded quantum dot), without any sign of significant elongation.
\label{fig:SEM} 
}
\end{suppfigure}

\textit{Quantum dot elongation} According to the atomistic million-atom many-body pseudopotential calculations by R. Singh et al., a small level of FSS ($\sim 3 \text{ - } 8\unit{\mu eV}$) appears upon $5\text{ - }15\%$ lateral elongation of pure InAs disk quantum dots ($h_\mathrm{D}=3.5 \unit{nm}$ and $D_\mathrm{D}=25 \unit{nm}$) embedded in [111]-oriented InP nanowires~\cite{Singh09}. For hexagonal quantum dots of comparable size, a similar range of FSS is expected. In our case of ternary In(As)P insertions with a large fraction of phosphorus ($\mathrm{InAs_{0.2}P_{0.8}}$) and $h_\mathrm{D}= 6\text{ - }8 \unit{nm}$, an even more pronounced elongation is required to induce this amount of FSS as the quantum dot confinement and piezoelectric potential are both weak and the orbitals are comparatively dilute. We do not observe any evident sign of lateral elongation in scanning electron microscopy (SEM) images showing the cross-sections of nanowire cores (see Fig.~\ref{fig:SEM} (b)). On average, a small elongation ratio ($<5\%$) is confirmed for the investigated nanowires, which is unable to induce a significant level of FSS ($>10~\mu$eV).

\begin{suppfigure}[h]
\centerline{\includegraphics[width=1\columnwidth]{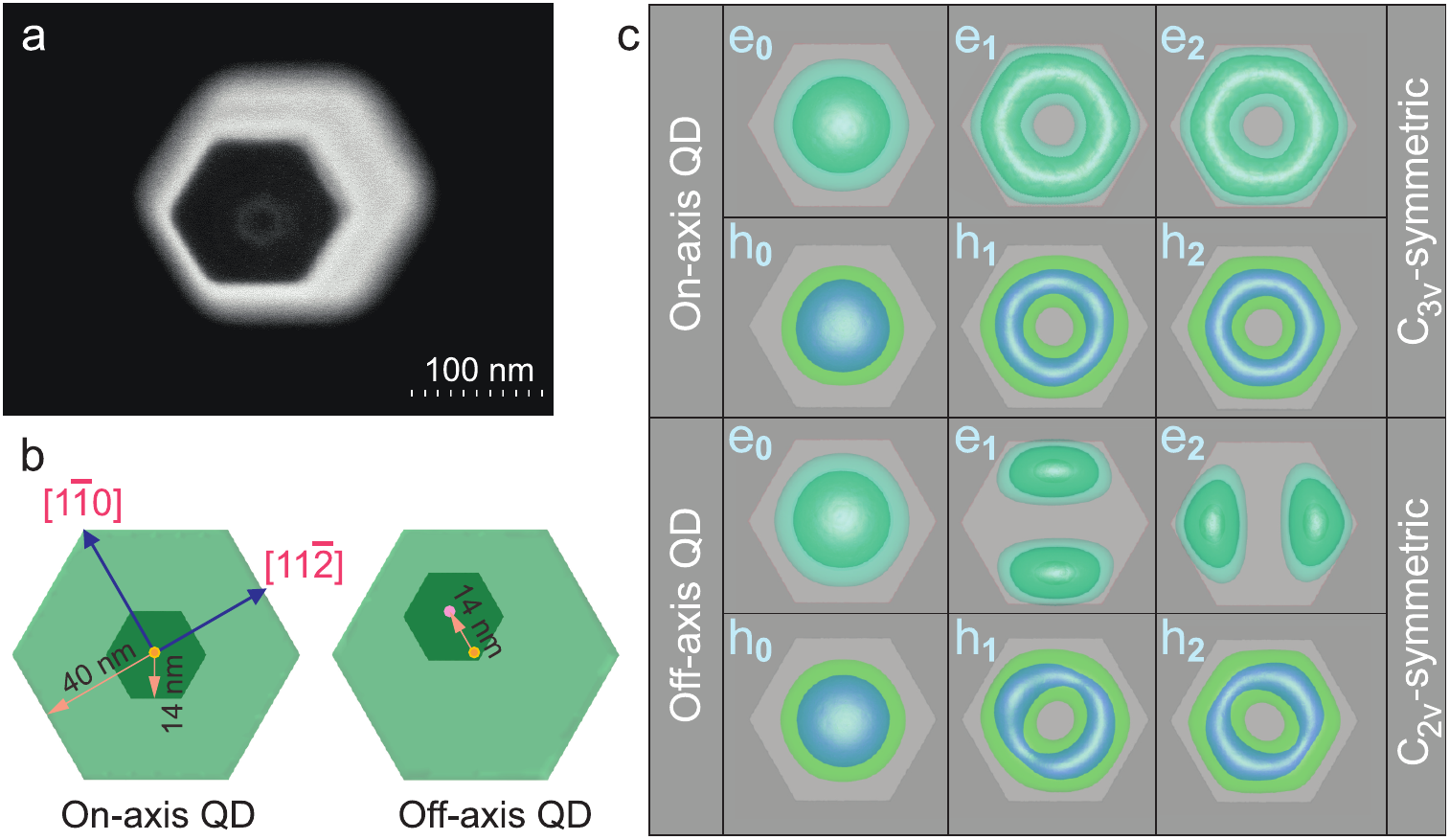}}
\caption{Wave function calculations for off-centered quantum dots. \textbf{a} Plan view SEM image of a clad nanowire depicting its cross section. The nanowire core is slightly displaced with respect to the center of cladding. Inset: plan view SEM image of a nanowire core showing a significant displacement off the center \textbf{b} Cross section of numerically modeled $\mathrm{InAs_{0.2}P_{0.8}/InP}$ NW quantum dots with hexagonal symmetry. Left, the quantum dot center and the cladding axis are aligned; Right, the quantum dot center is moved 14 nm along $[1\bar{1}0]$ direction with respect to the cladding axis. \textbf{c} The probability density of electrons (holes) in the $s$ shell, $e_{0}$ ($h_{0}$), and $p$ shells, $e_{1}$-$e_{2}$ ($h_{1}$-$h_{2}$) of NW quantum dots introduced in \textbf{b}. Only the quantum dot region is illustrated. Top: orbitals exhibit $C_{3v}$ symmetry when the nanowire core is located at the center of cladding. Bottom: single particle orbitals of misaligned NW quantum dot. The orbital symmetry is lowered down to $C_{2v}$ as a result of non-uniform strain field. \label{fig:on_off_axis} 
}
\end{suppfigure}

\textit{Off-center nanowire growth} Another source of FSS could be the dislocation of the quantum dot insertion with respect to the nanowire axis due to the displacement of the gold particle during the temperature ramp. Fig. \ref{fig:on_off_axis}(a) shows the plan view SEM image of a clad nanowire, where the nanowire core is misaligned with the axis of the cladding. A considerable displacement, as shown in the inset of Fig. \ref{fig:on_off_axis}(a), induces a non-uniform strain field and piezoelectric potential. We modeled the impact of a moderate displacement ($\Delta_\mathrm{D}=14\unit{nm}<0.2D_\mathrm{shell}$) on a hexagonal $\mathrm{InAs_{0.2}P_{0.8}}$ quantum dot with $h_\mathrm{D}=6 \unit{nm}$ and $D_\mathrm{D}=28 \unit{nm}$ embedded inside a 80 nm-thick InP nanowire (see Fig. \ref{fig:on_off_axis}(b)). The details of the modeling can be found elsewhere \cite{Khoshnegar12}. Fig. \ref{fig:on_off_axis}(c) shows the single particle orbitals of electrons ($e_{0}$-$e_{2}$) and holes ($h_{0}$-$h_{2}$) of a quantum dot with the above specifications. As expected, the single particle orbitals present a $C_{3v}$ symmetry once the quantum dot is located at the center of nanowire ($\Delta_\mathrm{D}=0$). The bottom panel of Fig. \ref{fig:on_off_axis}(c) shows the single particle states in a quantum dot 14 nm dislocated along the $[1\bar{1}0]$ direction. The $C_{2v}$ symmetry of the wave-function shows $p$-states ($e_{1}$-$e_{2}$ and $h_{1}$-$h_{2}$) which signifies the asymmetry of the underlying strain field. This low symmetry character is not quite visible in the ground state ($e_{0}$ and $h_{0}$) because a) the displacement is limited and the strain field partially relaxes within the cladding, and b) the compressive strain experienced by the $\mathrm{InAs_{0.2}P_{0.8}}$ insertion inside an InP matrix is weak. Therefore we expect that only a trivial part of the FSS, namely $<5~\mu\mathrm{eV}$, can be induced by the quantum dot dislocation in our investigated samples.     

\textit{Compositional inhomogeneity} The above discussions suggest that the FSS in our NW quantum dots mostly originates from the compositional anisotropy rather than the geometrical asymmetry. It is well known that any anisotropy along the main quantization axis can lead to a polarized piezoelectric field and FSS in SK dots \cite{Schliwa09,Zielinski13}. NW quantum dots are however immune to these types of axial anisotropy \cite{Khoshnegar12}, thus the FSS observed in our samples must be primarily induced by inhomogeneity of their ternary composition.

\end{document}